\documentclass[aps,amsfonts,showpacs]{revtex4}
\usepackage{amsmath}
\usepackage{graphicx}
\usepackage{cases}

\usepackage{natbib}
\usepackage{psfrag}
\usepackage{amssymb}
\usepackage{color}
\usepackage{hyperref}

\begin{document}

\def\be{\begin{equation}}
\def\ee{\end{equation}}
\def\bea{\begin{eqnarray}}
\def\eea{\end{eqnarray}}
\def\bml{\begin{mathletters}}
\def\eml{\end{mathletters}}
\def\b{\bullet}
\def\eqn#1{(~\ref{eq:#1}~)}
\def\no{\nonumber}
\def\av#1{{\langle  #1 \rangle}}
\def\m{{\rm{min}}}
\def\M{{\rm{max}}}
\newcommand{\ds}{\displaystyle}
\newcommand{\tc}{\textcolor}
\newcommand{\TODO}[2][To do: ]{{\textcolor{red}{\textbf{#1#2}}}}
%=============================================================================
%=============================================================================

\title{Length of adaptive walk on uncorrelated and correlated fitness landscapes}  
\author{Sarada Seetharaman and Kavita Jain}
\affiliation{Theoretical Sciences Unit, Jawaharlal Nehru Centre for Advanced Scientific Research, Jakkur P.O., Bangalore 560064, India}
\widetext
\date{\today}
\begin{abstract}
We consider the adaptation dynamics of an asexual population that
walks uphill on a rugged fitness landscape which is endowed with large
number of local fitness peaks.  
We work in a parameter regime where only those mutants that are
single mutation away are accessible, as a result of which the
population eventually gets trapped at a local 
fitness maximum and the adaptive walk terminates.  We study how the
number of adaptive steps taken by 
the population before reaching a local fitness peak depends on the
initial fitness of the population, the extreme value distribution of
the beneficial mutations and correlations amongst the fitnesses.  
 Assuming that the relative fitness difference between successive
 steps is small, we analytically calculate the average walk 
 length for both uncorrelated and correlated fitnesses in all extreme
 value domains for a given initial fitness. We present numerical
 results for the model where the 
 fitness differences can be large, and find that the walk
 length behavior differs from that in the former model in the
 Fr{\'e}chet domain of extreme value theory. We also discuss the 
relevance of our results to microbial experiments. 
\end{abstract}

\pacs{87.23.Kg, 02.50.Cw, 02.50.Ey}

\maketitle

\section{Introduction}
\label{intro}

Fitness is a quantitative measure of how successful an organism is in
a given environment - an organism with high fitness  has a better
chance of propagation within the population than those with lower
fitness. Fitness landscape defined as a map from genetic sequences to
(genotypic) fitness is a fundamental concept in the theory of
biological evolution \cite{Gavrilets:2004,Visser:2014}. But to construct a
fitness 
landscape for a microbe with merely hundred nucleotide sequence, one
needs to experimentally measure the fitness of $4^{100} \sim 10^{60}$
sequences which is not possible with the current technology. However
some empirical insights have been obtained regarding the qualitative 
nature of the fitness landscapes in the recent
years. Fitness have been measured for various microbes for a small
part (up to ten loci) of the genome which gives information about the
local topography of the fitness landscape \cite{Szendro:2013}. Large
scale fitness 
landscapes for about $70,000$ HIV sequences have also been
constructed \cite{Hinkley:2011}. A key result which has emerged from
these empirical studies is that the fitness 
landscapes are quite rugged {\it i.e.} they are endowed with moderately 
large number of local fitness peaks which are sequences fitter than
their nearest  neighbours. A related characteristic of such fitness
landscapes is that they are partially correlated \cite{Carneiro:2010,Miller:2011} which has the effect of reducing the number of local fitness peaks relative to a fully uncorrelated fitness landscape.  

Besides measuring fitness
landscapes directly, the dynamics of adaptation have also been
exploited to obtain  
insights into the structure of the underlying fitness landscape 
\cite{Burch:1999,Burch:2000,Miller:2011,Jain:2011a,Barrick:2013}.  
During adaptation, a population climbs
the fitness landscape, and in asexual populations, this process
occurs exclusively via beneficial mutations. However as advantageous mutations are rare, accounting for less 
 than $15\%$ of all mutations \cite{Eyrewalker:2007,Sniegowski:2010},
 experimental study of adaptation is  
 difficult. But 
 in recent times, it has been possible to track adaptive trajectories
 for several tens to thousands of generations, especially in microbial
 populations \cite{Barrick:2013}. It has been observed that initially 
 the population evolves quickly and then its fitness
 increases slowly towards different fitness plateau for the same initial
 fitness \cite{Korona:1994,Burch:2000,Fernandez:2007} thus supporting the
 conclusion that  fitness landscapes are rugged. On
 such fitness landscapes, while very large populations
 can reach the global fitness maximum quickly as they produce greater number
 of  mutants, smaller 
 populations stay trapped at a local fitness peak for a long
 time \cite{Iwasa:2004,Weinreich:2005b,Rozen:2008,Jain:2011a}. 
In recent experiments, the
 number of adaptive mutations that occur till the population reaches a 
 fitness plateau have been measured, and it has been found that the 
 population encounters a local fitness maximum within two
 \cite{Gifford:2011} to nine \cite{Rokyta:2009} substitutions.   

In this article, we address how the number of adaptive steps that a
population takes before it gets trapped at a local fitness peak
depends on the properties of the underlying fitness landscape. We
consider an asexual population in the {\it strong 
  selection-weak mutation} regime 
where the rate of mutations is low enough to produce only those
sequences that are single mutation away ({weak
  mutation}), and a  
mutation that confers a fitness benefit has a substantial probability
of spreading through the population while the neutral or
disadvantageous mutations get lost ({strong selection}). 
As a result of these assumptions, the entire population can be
represented by a single point in the 
fitness landscape, and performs an uphill \textit{adaptive walk} which
terminates once the population reaches a local fitness 
peak since a better fitness is at least two mutations away
\cite{Gillespie:1983,Gillespie:1991}. The properties of the adaptive
walk depend on the distribution of beneficial mutations which can be
found by appealing to the extreme value
theory (EVT) \cite{Gillespie:1991} since beneficial  
mutations are rare and therefore lie in the tail of the full fitness
distribution \cite{Eyrewalker:2007, Sniegowski:2010}. For independent
and identically 
distributed (i.i.d.) random variables, the EVT states that 
the distribution of the tails can 
belong to one of the three domains namely Weibull, Gumbel and Fr{\'e}chet
\cite{Sornette:2000}. 
%\tc{red} {Determining the domain to which the fitness of beneficial mutations belong is of interest because of its crucial role in the adaptation of the population, especially to harsh environments like exposure to antibiotics \cite{Bull:2005}. It is i
Interestingly, all the three extreme value domains have been observed in recent experiments. Although the exponential distribution for beneficial mutations belonging to the Gumbel domain has been most commonly seen \cite{Sanjuan:2004a, Rokyta:2006, Kassen:2006, Maclean:2010}, the fitness distribution of beneficial mutations belonging to Weibull \cite{Rokyta:2009, Bataillon:2011} and Fr{\'e}chet \cite{Schenk:2012} domains have also been observed. Here we study how the walk length depends on these three extreme value domains. Although, as
mentioned above, fitnesses are known to be correlated, much of the
previous work on the subject ignores correlations completely \cite{Rokyta:2006,Joyce:2008,Neidhart:2011,Jain:2011e}. Here we 
also investigate how correlations affect the number of adaptive
substitutions. Motivated by recent experiments on adaptive walks in
which a maladapted population starts at different fitness
\cite{Rokyta:2009,Gifford:2011,Sousa:2012}, we also analyse the
dependence of the walk length on the initial fitness.

According to the population genetics theory
\cite{Charlesworth:2010}, the probability that a beneficial mutation
will spread through the population increases with the relative fitness
difference between the mutant and the parent exponentially
fast towards unity. Using this probability function, we numerically
find that the adaptive walk is shortest in the Gumbel domain \cite{Seetharaman:2014}. However 
when the relative fitness difference is assumed to be small, this
probability is proportional to the relative fitness difference, and in 
this case, we find that the adaptive walks are shortest in the
Fr{\'e}chet domain and longest in the Weibull domain. Although the 
assumption of small fitness differences is biologically incorrect, 
especially in the Fr{\'e}chet domain, it is still interesting to
consider this model as 
it connects to other systems \cite{Neidhart:2011}, such as 
deterministically evolving populations \cite{Jain:2005,Sire:2006} and a 
gas of particles undergoing elastic collisions 
\cite{Bena:2007,Sabhapandit:2008}, and lends itself to analytical
calculations. We calculate the 
average walk length in all the three EVT domains for uncorrelated
fitnesses, and show that it depends logarithmically on the initial
rank of the population. Using results from the large deviation theory 
\cite{Touchette:2009}, we also obtain analytical expressions for the
walk length for correlated fitnesses, and find that the walk lasts
longer on correlated fitness landscapes as they have fewer local fitness peaks.

The article is organised as follows: In Sec.~\ref{model}, we describe
the model of fitness landscapes and adaptation dynamics 
employed here. We present a detailed analysis of the model which
assumes the 
fitness difference to be small on uncorrelated and correlated fitness
landscapes in 
Sec.~\ref{linear}, and then move on to describe our numerical results
for the full model that takes care of large relative fitness
differences in Sec.~\ref{full}. We finally conclude with a summary 
of our results, and their relevance to the experiments. 

%=============================================================================
%MODEL
%=============================================================================

\section{Model}
\label{model}

%-----------------------------------------------------------------------

\subsection{Fitness landscapes}

The adaptation model studied here is defined on a  
hypercube of dimension $L$ where each vertex corresponds to a 
binary sequence, as shown in Fig.~\ref{hs} for $L=4$. Each sequence is
assigned a fitness which is an i.i.d. random variable chosen from a
probability distribution. Experiments indicate that deleterious and neutral
mutations account for most of the weight 
in the fitness distribution, but a significant fraction comes from the
beneficial mutations as well \cite{Eyrewalker:2007}.  
Since the adaptation process is governed by these rare 
beneficial mutations, we need to consider the upper tail of the 
fitness distribution \cite{Gillespie:1991} which immediately suggests the
use of the extreme value theory and the related peak-over-thresholds
formulation described below \cite{Joyce:2008,Sornette:2000}. 

Consider the conditional cumulative distribution $P_{f_T}(f)$ for the
fitness $f$ chosen from the distribution ${\hat p}(f)$ above a large threshold $f_T$
which here refers to the wild type fitness. Formally, we have 
\bea
P_{f_T}(f)  &=& \textrm{Prob}(F -f_T < f | F > f_T) \\
&=& 1- \frac{{\hat
        q}(f+f_T)}{{\hat q}(f_T)} 
\eea
where ${\hat q}(f)=\int_f dg~{\hat p}(g)$. For large enough 
thresholds, the above cumulative distribution approaches the
Generalised Pareto Distribution (GPD) \cite{Sornette:2000}: 
\be 
P_{f_T}(f) \stackrel{\textrm{large}~f_T }{\longrightarrow} P(f,\tau)=1-\left[1+ 
      \frac{\kappa f}{\tau} \right]^{-1/\kappa}~,~ -\infty < \kappa <
\infty
\label{pot}
\ee
where $\tau$ is a scale factor and the shape parameter $\kappa$ can
take any real value. The limiting distribution with positive $\kappa$
corresponds to a power law distribution, and is obtained when ${\hat
  p}(f)$ itself decays algebraically. When $\kappa < 0$, the fitness
distribution (\ref{pot}) makes sense when $f < -\tau/\kappa$ and
therefore such a distribution is bounded above. This class of
distributions appears when ${\hat p}(f)$ is truncated. Finally, the
limit $\kappa \to 0$ gives an exponentially decaying function which is
obtained from unbounded distributions decaying faster than a power
law. For example, for the fitness distribution 
${\hat p}(f)=c f^{c-1}~e^{-f^c}, c > 0$, the conditional distribution
works out to be  
\bea
P_{f_T}(f) &=& 1- \frac{e^{-(f+f_T)^c}}{e^{-f_T^c}}\\
&\approx& 1-e^{-c f_T^{c-1} f} ~,~f_T \gg 1
\eea
Thus the tail of the conditional distribution is an
exponential, and the threshold fitness {$f_T$} and the
exponent {$c$} characterising the tail of the full distribution ${\hat
  p}(f)$ appear in the scale factor $\tau$. In summary, the
distribution $p(f,\tau)=d P(f,\tau)/df$ of {\it beneficial}  
mutations for i.i.d. fitnesses is a GPD, or in the language of the extreme
value theory, the distribution $p(f,\tau)$ can be of only three types viz.,
Weibull ($\kappa < 0$), Gumbel ($\kappa \to 0$) and Fr{\'e}chet
($\kappa > 0$) \cite{Sornette:2000}. A result from the extreme value theory
that we will need for subsequent discussion is regarding the largest
value of $L$ random variables, or in other words, the typical fitness
${\tilde f}$ of a local fitness peak. Since, in a set of $L$ random
variables, the number of fitnesses that exceed ${\tilde f}$ is one, we
have \cite{Sornette:2000}
\be
L\int_{\tilde f}^u df~p(f)=1
\label{fitrank}
\ee
where $u$ is the upper limit of the fitness distribution. 
This immediately yields 
\be
\tilde f =\tau \left(\dfrac{L^\kappa-1}{\kappa} \right)
\label{largest}
\ee
We will set $\tau=1$ in the rest of this article and denote the fitness distribution by $p(f)$. As we
are interested in adaptive changes, an uncorrelated fitness landscape
is generated by choosing fitnesses independently from $p(f)$. 

We introduce correlations between sequence fitnesses using a block model
\cite{Perelson:1995}, where a sequence of length $L$ is
assumed to be built of $B$ blocks, each of length $L_B=L/B$. The  $2^{L_B}$ fitnesses of each of the $B$ blocks is an i.i.d. random variable chosen from GPD 
and the fitness of the whole sequence is given by the average of the block
fitnesses. Fitness correlations arise because of common blocks between 
two sequences and can be changed by tuning the number of
blocks in the sequence. The two limits namely $B=1$ and $B=L$
produce fully uncorrelated and fully correlated fitness landscapes
respectively.  A measure of the ruggedness of a fitness
landscape is the number of local fitness peaks (defined as sequences
fitter than all of their one mutant neighbours) which decreases as the
fitness correlations increase \cite{Perelson:1995}. On correlated fitness landscapes, a local fitness peak is reached when the fitness of each  block is a local fitness maximum. As the probability of one of the $2^{L_B}$ sequences being a local fitness peak is $2^{L_B}/(L_B+1)$, the average number of local peaks on a correlated fitness landscape is given by $\left(\frac{2^{L_B}}{L_B+1}\right)^B$ \cite{Perelson:1995}. Thus on a fully  correlated fitness landscape, there is only one local (same
as global) fitness peak, whereas on fully uncorrelated fitness landscapes, 
there are on an average ${2^L}/(L+1)$ local fitness maxima.

%-----------------------------------------------------------------------

\subsection{Adaptive walk}
\label{aw_model}

\begin{figure}
\centering
\includegraphics[width=1.0 \textwidth,angle=0]{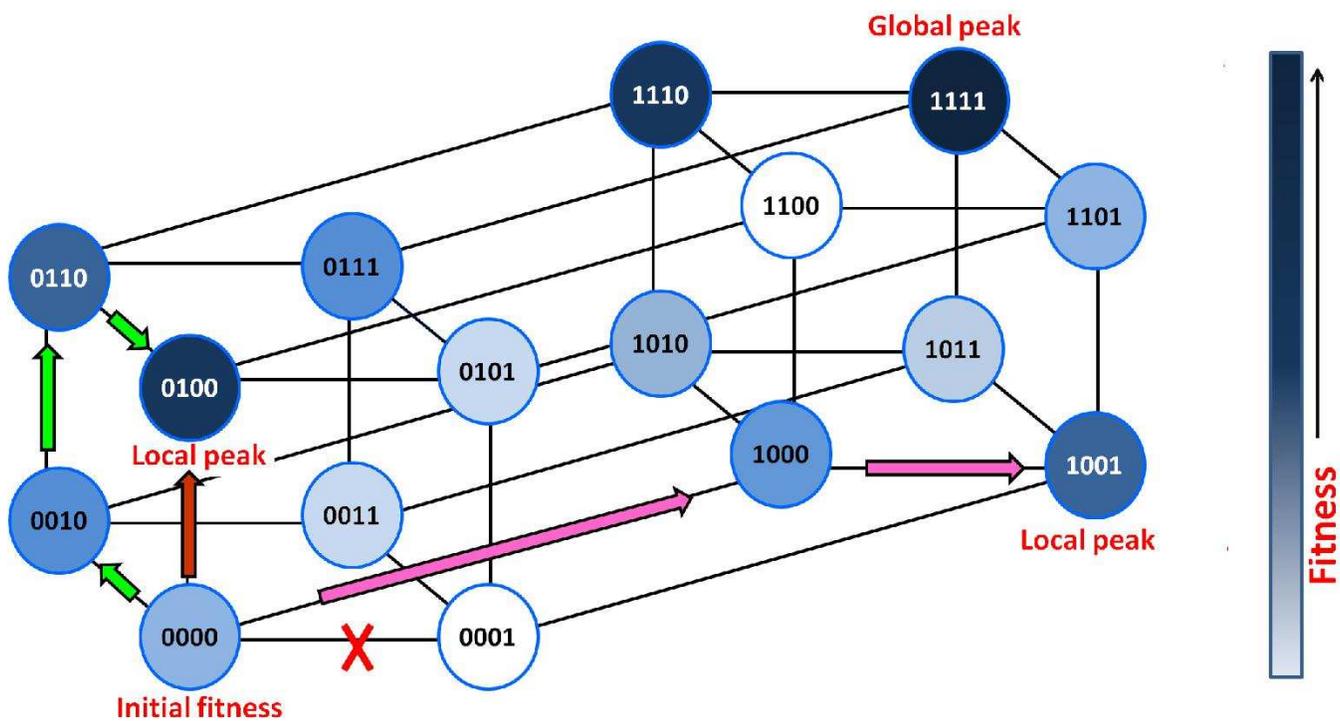} 
\caption{(Color online) Schematic representation of  
  adaptive walks in a $4$-dimensional sequence space, starting from the same initial sequence. The arrows represent
  the shift of the population from a sequence to a fitter sequence one
  mutation away (refer text for details).} 
\label{hs}
\end{figure}

We consider an asexual population initially localised at a
 sequence with fitness $f_0$ in the strong selection-weak
   mutation regime \cite{Gillespie:1991} in which only the beneficial
 mutations spread through the population and the mutation rates are
 small enough so 
that only those sequences that are one mutation away from the
 currently occupied sequence can be accessed. As illustrated 
in Fig.~\ref{hs} for sequence space of dimension $4$, starting from the
sequence $\{0000\}$, at the first step in the walk, the population has
three fitter  
neighbors viz. $\{0010 \}, \{0100\}$ and $\{1000 \}$, and it chooses one
of them according to a stochastic rule described below. After the first step is
taken, the population again scans its nearest neighbors and walks to a
fitter neighbor. This process repeats until a local fitness peak is
reached whereupon the adaptive walk terminates since the next beneficial
mutation is at least two mutations away which is not accessible in the weak 
mutation regime. The number of steps taken from the 
initial sequence to a local fitness peak is termed as 
the \textit{walk length}. In Fig.~\ref{hs}, two walks to the local fitness peak
$\{0100\}$ with length one and three are shown. Of course, an 
adaptive walk to a different local fitness peak (say, sequence
$\{1001\}$) is also possible. 

We now discuss the stochastic rules by which a nearest fitter sequence
may be chosen \cite{Orr:2003b}. Perhaps the simplest algorithm is the {\it
  greedy adaptive walk} (GAW) in which the fittest mutant is chosen at any
step in the walk. The average length ${\bar J}$ of the GAW has
been calculated by 
appealing to the theory of records, and for infinitely long
sequences, it turns out that \cite{Orr:2003b}
\be
{\bar J}_{GAW}=e-1 \approx 1.718
\label{GAW}
\ee
for any
fitness distribution.  In contrast, in the {\it random adaptive walk} (RAW), any
fitter one-mutant is equally likely to be chosen and in
this case, the average length of the walk diverges with the
sequence length. More precisely, the average walk length for zero initial fitness is given by
\cite{Flyvbjerg:1992} 
\be
{\bar J}_{RAW} \approx \ln L+ 1.099
\label{RAW}
\ee
and is independent of the choice of
the fitness distribution. Here we are interested in the biologically
relevant situation where, as one would intuitively expect, a mutant
which is much fitter than the wild type has a higher chance of
sweeping through the population than a mutant which is mildly
fitter. From the population genetics theory \cite{Charlesworth:2010},
it is known that in a 
large adapting asexual population, if $h$ is the fitness of the wild type and
$f > h$ is the fitness of the mutant,  
the probability that the mutant will take over the population is given
by 
\be
\pi(f,h) = 1-\textrm{exp}\left[-\frac{2 (f-h)}{h} \right]
\label{fixprob}
\ee
Thus, as in Fig.~\ref{hs}, when several of the $L$ nearest mutants are
beneficial, the population moves to one of them with a probability
proportional 
to $\pi$. The normalised transition probability is then 
given by \citep{Orr:2002,Jain:2011d,Seetharaman:2014}
\be
T(f \leftarrow h) =  \frac{1-e^{-\frac{2 (f-h)}{h}}}{\sum_{g>h}1-e^{-\frac{2 (g-h)}{h} }} \hspace{0.1in}{\text(full~model)}
\label{Tfull}
\ee
 The above
equation is clearly nonlinear in the fitnesses, and we have not been
able to obtain analytical results using the above transition
probability. However our previous work \cite{Seetharaman:2014} shows that
when $\kappa \leq 0$, the relative fitness difference $s=(f-h)/h$ between 
the mutations encountered is small, and we may therefore write $\pi(f,h)\approx 2 s$
\citep{Gillespie:1983,Gillespie:1991, Orr:2002, Jain:2011d} which gives us  
\be
T(f \leftarrow h)=\frac{f-h}{\sum_{g>h} 
  g-h} \hspace{0.1in}{\mathrm(linear~model)}
\label{linT}
\ee 
In this article, we shall refer to the model that uses
(\ref{Tfull}) as the {\it full model} and present numerical 
results for it in Sec.~\ref{full}. In the next section, we will study
in detail the {\it linear model} 
that employs (\ref{linT}) for all $\kappa$. The linear model is
interesting to study, not only because it is 
amenable to analysis, but also because the results obtained here appear in other
systems \cite{Neidhart:2011} viz. models of 
deterministically evolving populations \cite{Jain:2005,Sire:2006,Jain:2011c} 
and the Jepsen gas that  
describes a system of particles with random velocities undergoing
elastic collisions \cite{Bena:2007,Sabhapandit:2008}. Two  
variants of the linear model have been studied: while
\cite{Gillespie:1983} and \cite{Neidhart:2011} considered adaptation
in a single fixed neighborhood where the mutants are produced only at
the first step and the same are retained all through the walk, the
model studied in \cite{Rokyta:2006, Joyce:2008, Jain:2011d,
  Jain:2011e, Seetharaman:2014} assumes that  a 
new set of $L$ fitnesses (corresponding to 
the fitness of the one mutant neighbours) are generated at each step of the walk. Though
we shall use the latter model here,  it is interesting to note that
most results for the walk length are robust with respect to this assumption.  
%=============================================================================
%LINEAR MODEL
%=============================================================================
\section{Walk length in the linear model}
\label{linear}

%-----------------------------------------------------------------------

\subsection{On uncorrelated fitness landscapes}
\label{wl_uncorr}

For zero initial fitness, it has been shown that if the
mean ${\bar f}$ of the fitness distribution $p(f)$ is finite, the walk
length increases with the length of the 
sequence but remains constant otherwise
\cite{Jain:2011d,Jain:2011e}. To 
understand this transition at $\kappa=1$ above which ${\bar f}$ is
infinite, here we present a simple 
argument and refer the reader to \cite{Jain:2011e} for details. 
For $\kappa < 1$, as the transition probability (\ref{linT}) is nonzero
for finite fitness differences, the adaptive walk goes on
indefinitely for infinitely long sequence  
 or in other words, the adaptive walk length diverges with the sequence
 length $L$. 
A calculation for zero initial 
 fitness and large $L$ shows that the walk length cumulants 
increase logarithmically with the sequence length \citep{Jain:2011e}. In
particular, the mean walk length ${\bar J}$ increases as
\citep{Neidhart:2011,Jain:2011d,Jain:2011e} 
\be
{\bar J}(L|f_0=0) \approx \beta_\kappa \ln L
\label{beta2}
\ee
where 
\be
\beta_\kappa=\dfrac{1-\kappa}{2-\kappa} ~,~\kappa<1
\label{beta}
\ee
which shows that the walks are shorter for slowly decaying fitness
distributions. For $\kappa > 1$, as the mean of the fitness distribution 
is infinite, the normalisation sum in the denominator on the
right hand side (RHS) of (\ref{linT}) is dominated by the largest value ${\tilde f}$ amongst $L$ i.i.d. random variables
(refer (\ref{largest})). This implies that the transition occurs to one of the
highly fit sequences with fitness of order $\tilde f$. Since the
number of such sequences is of order unity, the walk terminates in a
few steps resulting in a constant walk length. 

\begin{figure}[t]
\centering
\includegraphics[width=0.8 \linewidth,angle=270]{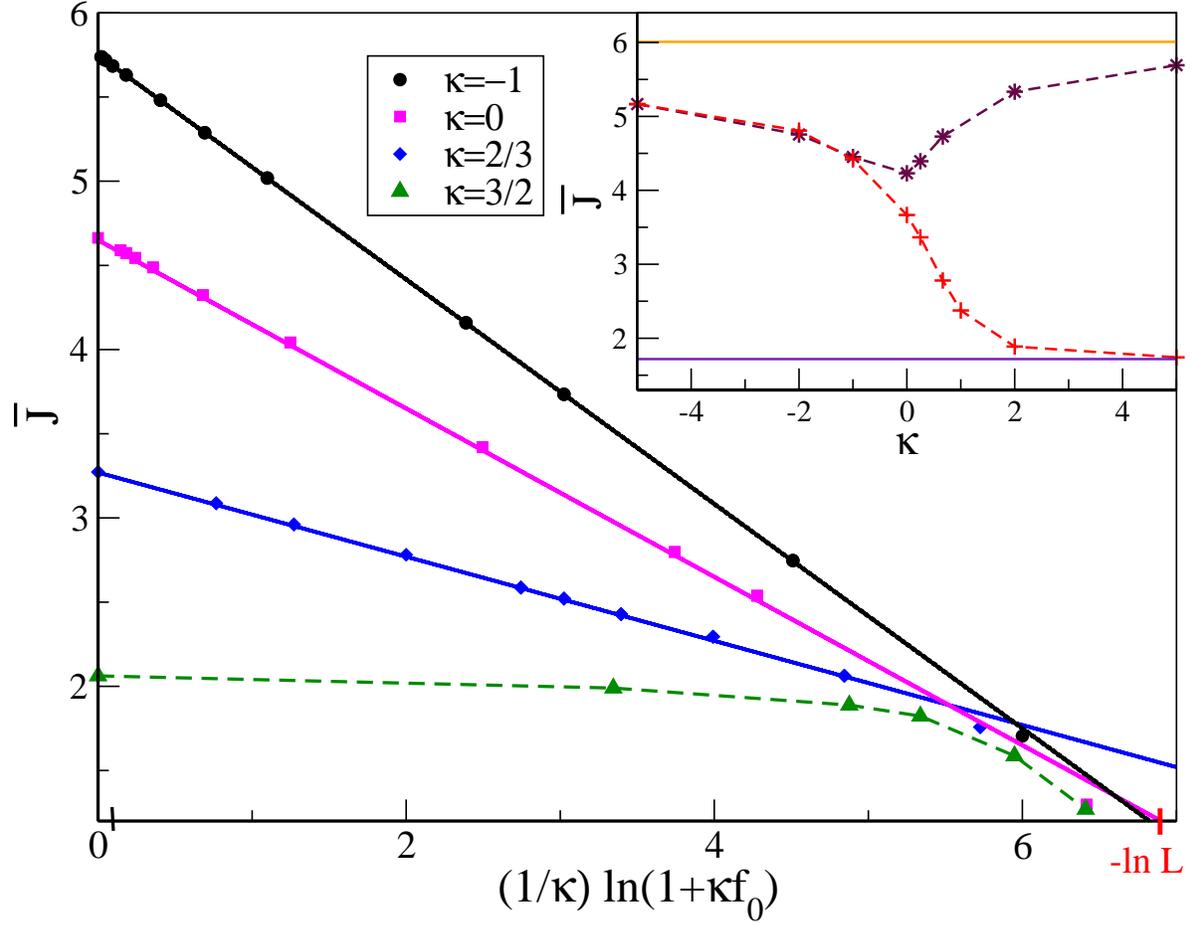} 
\caption{(Color online) Main: Variation of the average walk length with initial fitness
  in the linear model on uncorrelated fitness landscapes for
  various $\kappa$. The simulation points are for $L=1000$ and the
  lines are obtained from (\ref{wlfit}) for all $\kappa<1$, while the
  one for $\kappa=3/2$ is a guide to the
  eye. Inset: Comparison of the average walk length in the full model
  ($\ast$) and the linear model (+) for $({1}/{\kappa}) \ln(1+\kappa
  f_0)=2$. The solid line shows the walk length expressions
  (\ref{GAW}) (bottom) and (\ref{RAW}) (top) for the greedy 
  walk and the random adaptive  walk respectively.}       
\label{trans_fig}
\end{figure}

As shown in Fig.~\ref{trans_fig}, a similar transition is 
seen at $\kappa=1$ when the sequence length is kept fixed and the
initial fitness is varied. We now generalise the calculation in
\cite{Jain:2011e}  for zero 
initial fitness to find how the average walk length %(\ref{avgdefn}) 
changes 
with the initial fitness when $\kappa < 1$. Since the mean of the
fitness distribution is finite when $\kappa$ is below unity, for long
sequences, we can write (\ref{linT}) as \cite{Jain:2011d}
\be
T(f \leftarrow h)=\frac{p(f)~(f-h)}{\int_h^u dg~p(g)~(
  g-h)} 
\label{linT1}
\ee 
using which we will calculate the walk length as detailed below. 

An adaptive walk will stop at step $J$ if all the $L$ neighbouring
sequences have a fitness lower than that of the 
currently occupied sequence. Thus if $Q_J(L|f_0)$ is the probability that
the adaptive 
walk of a sequence of length $L$ lasts exactly $J$ steps, we can write
\citep{Jain:2011d}   
\be
Q_J(L|f_0)=\int_{f_0}^u df~ q^L(f)~{\cal P}_J(f|f_0)
\label{Qj}
\ee 
where ${\cal P}_J(f|f_0)$ is the probability distribution of the
fitness $f$ at the $J$th step, given the initial fitness $f_0$ which
satisfies (\ref{main}), and 
$q(f)$ is the cumulative probability of having a fitness lower than $f$ which is
given by   
\be
 q(f)=\int_0^f dg~p(g)=1-(1+\kappa f)^{-1/\kappa}
\label{qf}
\ee
%The quantity of our interest viz., the average walk length
%is given by ${\bar J}(L|f_0)=\sum_{J=0}^{\infty} J ~Q_J(L|f_0)$. 
%% For $\kappa<1$ the average walk length can be calculated using the transition probability (\ref{linT}) which for large $L$ can be written as \cite{Jain:2011d}
%% \be
%% T(f \leftarrow h)=\frac{p(f)~(f-h)}{\int_h^u dg~p(g)~(
%%   g-h)} 
%% \label{linT1}
%% \ee 
For the transition
probability (\ref{linT1}), the integral equation 
(\ref{main}) for the distribution ${\cal P}_J(f|f_0)$ appearing in
(\ref{Qj}) can be 
recast as a second order differential equation for the distribution
$P_J(f|f_0)$ defined through ${\cal 
  P}_J(f|f_0)=p(f) P_J(f|f_0)$, and is given by \citep{Jain:2011d}   
\be
{P}^{''}_{J+1}(f|f_0)=\frac{p(f) (1-q^L(f))}{\int_f^u
 dg~ (g-f)~p(g)} { P}_J(f|f_0)~,~ J \geq 1
\label{final}
\ee
where the prime refers to a derivative with respect to (w.r.t.)
$f$. Although we are unable to analyse 
(\ref{final}) when $L$ is finite, as explained below, it is possible to
extract useful information from it when the sequence is
infinitely long and using the fact that for a finite 
sequence, there is a characteristic fitness scale ${\tilde f}$ given
by (\ref{largest}).

We first introduce the generating function $G(x,f)=\sum_{J=1}^\infty
P_J(f)~x^J, x < 1$ which, due to (\ref{final}), 
obeys the following differential equation: 
\bea
 G''(x,f) 
= \frac{x (1-\kappa)(1-q^L(f))}{(1+\kappa f)^2} G(x,f)
\label{G}
\eea
and is subject to the initial conditions (\ref{bg1}) and (\ref{bg2}). 
In the above equation, the cumulative probability $q^L(f)$ of the
maximum value distribution is a smoothly varying function that
increases from zero to one, as the fitness $f$ increases and belongs to
one of the three EVT domains. For the cumulative fitness distribution
(\ref{pot}), we find that for large $L$ \citep{Sornette:2000} 
\begin{subnumcases}
{\label{qLdef} q^L(f) \approx e^{-(\frac{1+\kappa f}{1+\kappa {\tilde
  f}})^{-\frac{1}{\kappa}}}=}
e^{-z^{-\frac{1}{\kappa}}}~~~,~\kappa < 0 ~~(\textrm{Weibull})\\
e^{-e^{-z}}~~~~,~\kappa \rightarrow 0 ~~(\textrm{Gumbel})\\
e^{-z^{-\frac{1}{\kappa}}}~~~,~\kappa > 0 ~~(\textrm{Fr{\'e}chet})
\end{subnumcases}
where
\begin{subnumcases}
{\label{zdef} z(f)=} f-{\tilde f} &~,~$\kappa \rightarrow 0$\label{zdef1}\\
(1+\kappa f) (1+\kappa {\tilde f})^{-1} &~,~$\kappa \neq 0$
\label{zdef2}
\end{subnumcases}
It is useful to consider (\ref{G}) as a function
of $z$ defined above. If $\tilde z \equiv z({\tilde f})$, the
general solution of the differential equation (\ref{G}) may be 
written as 
\begin{subnumcases} 
{G(x,z)=} a_1 g_1(x,z)+a_2 g_2(x,z) &, $z < {\tilde z}$ \label{Gless}\\
b_1 h_1(x,z)+b_2 h_2(x,z) &, $z > {\tilde z}$
\end{subnumcases}
where $g_i,h_i$ satisfy (\ref{G}), and 
the constants $a_1, a_2$ are determined in Appendix~\ref{Agenfn} using
the initial conditions at $z_0 \equiv z(f_0) < {\tilde z}$. The
other constants of integration $b_1,b_2$ can be found by matching the 
solution $G(x,z)$ and its 
first derivative (w.r.t. $z$) at $z={\tilde z}$. 
Noting that ${\tilde z}$ is constant in $L$ and $f_0$ but
$z_0$ depends on them, 
%on solving the above set of simultaneous linear equations, 
we find that the constants $b_1, b_2$ are of the form 
\be
b_i= b_{i1}(x) a_1(z_0)+ b_{i2}(x) a_2(z_0) ~,~i=1,2
\label{bdef}
\ee

To find the properties of the walk length, we next define a generating
function $H$ for the walk length distribution (\ref{Qj}) as
\bea
H(x,L) &=& \sum_{J=1}^\infty Q_J(L|f_0) ~x^J \\
&=&\int_{z(f_0)}^{z(u)} dz~p(z)~\frac{dz}{df}~q^L(z)~G(x,z) 
\eea
On approximating $q^L(z)$ for $z<{\tilde z}$ by zero, we get  
\be
H(x,L)\approx \int_{\tilde{z}}^{z(u)}dz~p(z)~\frac{dz}{df}~q^L_>(z)~G_>(x,z)
\label{Happrox}
\ee
where the subscript $>$ is used to denote the quantities when $z>\tilde
z$. Using (\ref{zdef}) and (\ref{bdef}), we can
extract the $z_0$-dependence of the generating function and find that 
\begin{subnumcases}
{\label{cum} H(x,L)=} 
a_1(z_0) R_1(x) +a_2(z_0) R_2(x) ~~~~~~~~~~~~~,~\kappa \rightarrow 0
\label{Hcases1}\\
\frac{\kappa}{1+\kappa {\tilde f}} (a_1(z_0) R_1(x) +a_2(z_0) R_2(x))
~,~\kappa \neq 0 
\label{Hcases2}
\end{subnumcases}
where 
\be
R_i(x)=\int_{\tilde{z}}^{z(u)} dz~p(z)~q^L_>(z)~\sum_{j=1}^2 b_{ji}~h_j(x,z)~
\label{T}
\ee
is independent of $L$ and $f_0$. 
Furthermore, from the  explicit expressions for $a_1$ and
$a_2$ given in Appendix~\ref{Agenfn}, we 
see that $a_2$ decays more rapidly with $L$ than $a_1$, and therefore
we may neglect the second term on the RHS of (\ref{Hcases1}) and
(\ref{Hcases2}) for large $L$.  Since the $n${th} cumulant $\mu_n$ of
the walk length is given by \citep{Sornette:2000}
\begin{equation}
 \mu_n(L)=\dfrac{d^n \ln H}{d X^n}\bigg |_{X=0}
\end{equation}
where $X=\ln x$, to leading order in $L$, we finally obtain
\begin{subnumcases}
{\label{cum1} \mu_n(L)\approx}
(\ln L -f_0)\dfrac{d^n}{d X ^n} e^{X/2}\bigg|_{X=0} &,~$\kappa\rightarrow 0$\\
 \dfrac {1}{2 \kappa} \ln \left(\frac{L^\kappa}{1+\kappa f_0}
 \right)~\dfrac{d^n}{d X^n}\sqrt{\kappa^2+4 e^{X}
   (1-\kappa)}\bigg|_{X=0} &,~$\kappa\neq 0$
\end{subnumcases}

Setting $n=1$ in our final result (\ref{cum1}), we find the average
walk length to be 
\be
{\bar J}(L|f_0)  = \beta_\kappa \left(\ln L -\dfrac{1}{\kappa}\ln (1+\kappa f_0)\right) +c_\kappa
\label{wlfit}
\ee
where $\beta_\kappa$ is given by (\ref{beta}) and the constant $c_\kappa$ in which
the subleading corrections in $L$ are subsumed is determined numerically. We
check that the results of  \cite{Jain:2011d} and \cite{Jain:2011e}
for $f_0=0$ are reproduced from the above equation. \tc{black}{We also note that
since the {\it typical} rank $m$ of a fitness (with the fittest ranked one) is given
by \citep{Sornette:2000} 
\be
m= \frac{L}{(1+\kappa f_0)^{\frac{1}{\kappa}}}=\left(\frac{1+\kappa {\tilde f}}{1+\kappa f_0} \right)^{\frac{1}{\kappa}}
\label{rank}
\ee
our result (\ref{wlfit}) gives ${\bar J}=\beta_\kappa \ln m+c_\kappa$. Thus the effect of nonzero initial fitness is to replace the sequence length $L$ in (\ref{beta2}) for zero initial fitness (where all the mutants are fitter) by the average number of mutants present at the beginning of the walk. The logarithmic dependence of the walk length on the initial rank has been obtained in \cite{Gillespie:1983,Neidhart:2011} using a model in which both  
the initial rank $m$ and the mutational neighborhood are {\it fixed}. Here instead the initial fitness is fixed, but the initial rank is a random variable and a new suite of mutants is generated at every step in the walk. The fact that the same basic result is obtained in the deterministic and stochastic model shows that the stochastic effects are rather unimportant on an average as noted in previous works as well \cite{Rokyta:2006,Neidhart:2011}.}
% For zero initial fitness, since the initial rank is $L$,On comparing this result with (\ref{}) for zero initial fitness, we see that the effect of initial fitness is to effectively reduce the sequence length This expression is in line with the results of previous works obtained numerically on fitness landscapes that change with every step of the adaptive walk \cite{Rokyta:2006} and the results obtained from a model
% with fixed mutational neighborhood \cite{Neidhart:2011}. The intuitive understanding of this is that the population sees only those sequences fitter than the starting fitness and the consistency in the results obtained from different models indicates that the fluctuation in the rank of the initial fitness is small enough to not affect the average results. } 

Our numerical results for the average walk length on uncorrelated fitness
landscapes are compared 
  with (\ref{wlfit}) in Fig.~\ref{trans_fig} where the numerical fits
  for  
  constants $c_\kappa$ for $\kappa=-1$, $0$ and $2/3$ are $1.15$,
  $1.21$ and $1.55$ respectively. We see a good match between the simulation
data and (\ref{wlfit}) except when the initial fitnesses are close
to the local fitness optimum where the simulation data
  lies below the theoretical results. This discrepancy may be
  due to the fact that the approximation $q(f)=0$ is good for
  fitnesses far below the local fitness 
  peak, while we have used it for all $f < {\tilde f}$ to arrive at
  (\ref{Happrox}).

%-----------------------------------------------------------------------

\subsection{On correlated fitness landscapes}
\label{corr}

In the above discussion, we have assumed that the sequence fitnesses are
uncorrelated. We now discuss how the walk length changes when
correlated fitnesses generated using a block model (described in the
last section) are considered. If a sequence is divided into
  $B$ blocks and the initial fitness of the $b${th} block is
  $f_0^{(b)}$, the initial fitness of the whole sequence is given by 
\be
f_0=\dfrac{1}{B}\sum _{b=1}^Bf_0^{(b)}
\label{f0avg}
\ee
Since the block fitnesses evolve independently, the average walk length is the sum of the mutations accumulated by each block \cite{Perelson:1995,Jain:2011d}.
%A straightforward generalisation of the calculation in
%\cite{Jain:2011d} shows that 
Thus the
average walk length ${\bar J}_B$ for a 
sequence composed of $B$ blocks is given by 
\be
{\bar J}_B (L|f_0) =\sum_{b=1}^{B} {\bar J}(L_B|f_0^{(b)})
\label{anyB}
\ee
where ${\bar J}(L_B|f_0^{(b)})$ is the average walk length for a
sequence of length $L_B$ with initial fitness $f_0^{(b)}$ on
uncorrelated fitness landscapes. 
%\tc{red}{An earlier work \cite{Perelson:1995} showed that the average walk length of the whole sequence is the sum of the  independent contribution of each block since the probability of the walk length lasting $\bar{J}$ steps is obtained from the product of the probabilities of each block taking only $j\le \bar{J}$ steps such that $\sum_{i}^B={\bar J}$. The equation we have obtained in (\ref{anyB}) is consistent with the observation of the previous work mentioned above. } 
In the
simplest situation where the initial fitness $f_0^{(b)}$ of each block is same, we immediately have  \citep{Perelson:1995,Jain:2011d} 
\be
{\bar J}_B (L|f_0)= B {\bar J} (L_B|f_0) 
\label{idblk}
\ee

However if the block fitnesses are random variables that satisfy
(\ref{f0avg}), an average over the joint distribution $P_B(\{f_0^{(b)}
\})$ of block fitnesses is also required. We thus have  
\bea
{\bar J}_B (L|f_0) = \int_0^{u} df_0^{(1)} ...\int_0^{u} df_0^{(B)} P_B(\{f_0^{(b)}\}) 
\sum _{b=1}^B {\bar J}(L_B|f_0^{(b)})   
\label{randomic}
\eea
Since the block fitnesses are i.i.d. random variables subject to the
constraint (\ref{f0avg}), the distribution of block fitnesses can be
written as  
\be
P_B(\{f_0^{(b)}\}) = \frac{\prod_{b=1}^B p(f_0^{(b)})}{{\cal N}_B(B
  f_0)} ~\delta(B f_0-\sum _{i=1}^B f_0^{(i)})
\label{PB} 
\ee
where the normalisation constant ${\cal N}_B(X)$ is the distribution
of the sum of $B$ random variables given by    
\bea
{\cal N}_B(X) &=& \int_0^{u} df_0^{(1)}...\int_0^{u} df_0^{(B)}
\prod_{b=1}^B p(f_0^{(b)}) ~\delta \left (X-\sum _{i=1}^B
f_0^{(i)}\right) \label{norm} \\ 
&=& \int_0^{X} df ~p(f) ~{\cal N}_{B-1}(X-f) 
\eea
with ${\cal N}_0(f)=\delta(f)$. 
Thus we can express 
the average walk length as  
\be
{\bar J}_B (L|f_0)=B (\beta_\kappa \ln L_B + c_\kappa)
-\frac{\beta_\kappa B}{\kappa} \frac{ \int_{l_1}^{l_2} df ~p(f)  \ln (1+\kappa f)~
{\cal N}_{B-1}(B f_0- f)}{{\cal N}_B(B f_0)} 
\label{Jsimple}
\ee
where the integration limits are $l_1=0, l_2=B f_0$ in the Gumbel and Fr{\'e}chet
domains. In the Weibull domain, three cases arise: (i) if $B f_0 < u$,
the limits are $l_1=0, l_2=B f_0$, (ii) if $u < B f_0 < (B-1) u$, we have
$l_1=0, l_2=u$ and (iii) if $(B-1) u < B f_0 < B u$, the limits are
$l_1=B f_0-(B-1) u, l_2=u$.

%-----------------------------------------------------------------------

\subsubsection{Exactly solvable case}
\label{exp_corr}

For exponentially distributed fitnesses, the
distribution ${\cal N}_B(X)$ in (\ref{norm}) is known exactly 
 to be \citep{Feller1:2000}
\be
{\cal N}_B(X)= e^{-X}\frac{X^{B-1}}{(B-1)!}
\label{expo_exact}
\ee
Taking the limit $\kappa \to 0$ in
(\ref{Jsimple}), we find the average walk length as
\bea
{\bar J}_B (L|f_0) &=& B (\beta_0\ln L_B+c_0)-B \beta_0
\frac{\int_0^{B f_0} df ~e^{-f} ~f ~{\cal N}_{B-1}(B f_0-f)}{{\cal
    N}_B(B f_0)} \label{correl_exp2}\\ 
&=& B (\beta_0\ln L_B+c_0)-B \beta_0 \frac{e^{-B f_0} (B f_0)^B}{B
  ! ~{\cal N}_B(B f_0)} \\
&=& B {\bar J}(L_B|f_0)
\label{correl_exp}
\eea
which is the same as that in the case where each block fitness is $f_0$
(refer (\ref{idblk})).

%-----------------------------------------------------------------------

\subsubsection{Weakly correlated fitnesses}
\label{w_corr}

For $\kappa \neq 0$, it appears difficult to obtain exact expressions
for the walk length for correlated fitnesses. The case of two
independent blocks ($B=2$) presents the simplest model for
correlated fitnesses, and we discuss this here. The distribution ${\cal N}_2(X)$ of two random variables
is given by  
\begin{subnumcases}
{{\cal N}_2(X)= \label{N2}}
\int_0^X dg~p(g)~ p(X-g) ~,~X < u \label{N21}\\
\int_{X-u}^u dg~p(g)~ p(X-g) ~,~X > u \label{N22}
\end{subnumcases}
%\tc{red}{While (\ref{N21}) holds for all $f_0$ in the Gumbel and Fr{\'e}chet domains and for $f_0<u/2$ in the Weibull domain, (\ref{N22}) applies only in the Weibull domain for $f_0>u/2$.} 
For $2 f_0 < u$, using (\ref{N21}) in the expression (\ref{Jsimple}), 
we get 
\bea
\frac{{\bar J}_2 (L|f_0)}{B} &=& \beta_\kappa \ln L_B + c_\kappa 
-\frac{\beta_\kappa }{\kappa} \frac{ \int_{0}^{2 f_0} df ~p(f)  \ln
  (1+\kappa f) ~p(2 f_0- f)}{\int_{0}^{2 f_0} df ~p(f) ~ p(2 f_0- f)} \\
&=& \beta_\kappa \ln L_B + c_\kappa -\frac{\beta_\kappa }{\kappa} \ln(1+\kappa f_0)
+ \frac{\beta_\kappa}{2 \kappa} {\cal I}_{\kappa}(w_0) 
\label{f0ltu}
\eea
where the integral 
\be
{\cal I}_\kappa (w_0)=\frac{\int_{1}^{w_0} dz~\ln z~
  z^{\frac{1-\kappa}{\kappa}} (1-z^{-1})^{-1/2}}{\int_{1}^{w_0} dz~
  z^{\frac{1-\kappa}{\kappa}} (1-z^{-1})^{-1/2}}
\ee
with $w_0=(1+\kappa f_0)^2/(1+2 \kappa f_0)$. Note that for large initial
fitnesses $f_0 \sim u/2$, the function $w_0 \gg 1$. 

\noindent{\it Fr{\'e}chet class:} For positive $\kappa$ and large $f_0$, an
approximate expression for the integral ${\cal I}_\kappa(w_0)$ can be obtained
after an integration by parts, and we get  
\bea
\frac{{\bar J}_2 (L|f_0)}{B} &\approx& {\bar J}(L_B|f_0)+ \frac{\beta_\kappa}{2
  \kappa}(\ln w_0-\kappa) \\
&\approx& \beta_\kappa \ln L_B + c_\kappa -\frac{\beta_\kappa }{2 \kappa} (\ln(\kappa f_0)+\ln
2+\kappa)  
\label{B2Fre}
\eea

\noindent{\it Weibull class:} The integral ${\cal I}_\kappa(w_0)$ can be
calculated exactly for uniformly distributed fitnesses and is given by  
\bea
{\cal I}_{-1}(w_0) &=& 2 + \ln w_0 - 2 \sqrt{\frac{w_0}{w_0-1}} \sinh^{-1}
(\sqrt{w_0-1}) \\
&\approx& 2 (1-\ln 2) - \frac{\ln w_0}{2 w_0}  ~,~ w_0 \gg 1
\eea
For arbitrary negative $\kappa$, we note that the integral ${\cal
  I}_\kappa(w_0)$ is finite when $w_0 \to \infty$ and can be written
in terms of the harmonic number $H_n=n \sum_{i=1}^\infty {(i (n+i))}^{-1}$ \cite{Knuth:1997}. An
integration by parts then yields  
\be
{\cal I}_\kappa(w_0) \approx H_{-\frac{1}{\kappa}-\frac{1}{2}}-H_{-\frac{1}{\kappa}-1}+
 \frac{\kappa \Gamma(\frac{1}{2}-\frac{1}{\kappa})}{\sqrt{\pi} \Gamma(-\frac{1}{\kappa})} ~\frac{\ln
   w_0}{w_0^{1/|\kappa|}} 
\ee
which matches the result for $\kappa=-1$ as $H_{1/2}=2 -\ln 4$. 
Ignoring the last term on the RHS of the above equation which decays
with $f_0$, we find that the average walk length can be written as 
\be
\frac{{\bar J}_2 (L|f_0)}{B}={\bar J}(L_B|f_0)+ \frac{\beta_\kappa}{2 \kappa}
(H_{-\frac{1}{\kappa}-\frac{1}{2}}-H_{-\frac{1}{\kappa}-1})~,~f_0
\lesssim u/2
\label{B2W1}
\ee
For $f_0 > u/2$ where ${\cal N}_2(X)$ is given by (\ref{N22}), the integrals can be done exactly and we have 
\bea
\frac{{\bar J}_2 (L|f_0)}{B}&=&\beta_\kappa \ln L_B + c_\kappa-\frac{\beta_\kappa}{\kappa}
\frac{\int_{-1}^{1} dh~
(1-h^2)^{-\frac{1+\kappa}{\kappa}} ~(\ln(1+\kappa
  f_0)+\ln(1+h))}{\int_{-1}^{1} dh~
  (1-h^2)^{-\frac{1+\kappa}{\kappa}}} \\
&=& \beta_\kappa \ln L_B -\frac{\beta_\kappa}{\kappa} \ln(1+\kappa f_0)+
c_\kappa+\frac{\beta_\kappa }{2 \kappa}
(H_{-\frac{1}{\kappa}-\frac{1}{2}}-H_{-\frac{1}{\kappa}-1})  \label{B2up}
\\
&=& {\bar J}(L_B|f_0)+\frac{\beta_\kappa }{2 \kappa}
(H_{-\frac{1}{\kappa}-\frac{1}{2}}-H_{-\frac{1}{\kappa}-1}) ~,~f_0 > u/2 \label{B2W2}
\eea

\tc{black}{For bounded distributions, although the walk length is continuous at initial fitness equal to $u/2$, it is interesting to note that it is not differentiable. For uniformly distributed fitnesses where exact expressions for the walk length can be calculated, the average walk length obtained from (\ref{f0ltu}) and (\ref{B2W2}) is found to be the same at $f_0=1/2$. The first derivative of the walk length (with respect to $f_0$) is given by 
 \begin{subnumcases}
{\frac{d{\bar J}_2}{df_0}=}
%(2 beta ((-1 + f0) f0 + 
%   Sqrt[1 - 2 f0] Sqrt[-((-1 + f0)^2/(-1 + 2 f0))]
%     ArcSinh[f0/Sqrt[1 - 2 f0]]))/((-1 + f0) f0^2)
\frac{2 \beta_{-1}}{f_0^2} \left[f_0+ \frac{1}{2} \ln(1-2 f_0) \right] & ~,~  $f_0 < 1/2$ \label{differen1} \\
-\frac{2 \beta_{-1}}{1-f_0}  & ~,~  $f_0 > 1/2$ \label{differen2}
 \end{subnumcases}
From the above equation, we see that while the derivative at $f_0=1/2$ obtained from (\ref{differen1}) is undefined, the expression (\ref{differen2}) yields a finite constant. For general $\kappa < 0$, the derivative of the walk length calculated using (\ref{B2W2}) is seen to be finite, while it diverges when (\ref{f0ltu}) is used.}

%-----------------------------------------------------------------------

\subsubsection{On strongly correlated fitness landscapes}
\label{wl_corr_st}

We now turn to the situation when the block number $B \gg 1$. To
calculate the integral in (\ref{Jsimple}), let us first consider the
integrand   
\be
{\cal F}(f)= p(f)  \ln (1+\kappa f)^{1/\kappa} ~{\cal N}_{B-1}(B f_0- f)
\ee
The first two factors on the RHS are obviously
independent of $B$ and $f_0$. However for all 
$\kappa < 1$ where the fitness distribution has a finite mean ${\bar f}$,
the last factor peaks about the mean $B
(f_0-{\bar f})$ of the sum distribution which increases with both $B$ and
$f_0$. Then for large enough $B$ and $f_0$, the integrand 
${\cal F}(f)$ gets a contribution from the {\it lower
  tail} of the sum distribution instead of the region around its mean.  
The behavior of the tail of the sum distribution can be
obtained by applying a large deviation principle 
 if the fitness distribution possesses all finite
 moments as is the case for $\kappa \leq 0$. However for power law distributions with $\kappa > 0$,
 the $(1/\kappa)$-th and higher 
moments diverge and the large deviation principle is not applicable,
and in this case, we use the result  
that the sum distribution decays as the fitness distribution itself
\citep{Sornette:2000,Touchette:2009}. The fact that the central limit
theorem for the 
sum distribution does not capture the correct behavior of the integral
under question is illustrated in Appendix~\ref{gauss} for
exponentially distributed fitnesses.

To calculate the walk length using the large deviation theory, 
we first consider a normalised distribution with support on the
interval $[0,u]$ defined as  
\be
g(t)=\kappa (\alpha-1) (1+ \kappa t)^{-\alpha}
\label{gt}
\ee
where $\alpha < 1, u=-1/\kappa$ for $\kappa < 0$  and $\alpha > 1, u=
\infty$ when $\kappa \geq 0$.  
Then  the distribution of the sum of $B$ i.i.d. random
variables chosen 
from $g(t)$ is given by  
\be
I_B(X;\alpha)=\int_0^{u}dt^{(1)} ...\int_0^{u} dt^{(B)}  \prod_{j=1}^B g(t^{(j)})~
\delta \left (X-\sum _{i=1}^B t^{(i)}\right)
\label{sumI}
\ee 
Differentiating on both sides w.r.t. $\alpha$, we get
\be
\frac{\partial I_B(X;\alpha)}{\partial \alpha}= \frac{B I_B(X;\alpha)}{\alpha-1} -B \int_{0}^{\min(X,u)} dt ~g(t)  \ln (1+\kappa t) ~
I_{B-1}(X- t;\alpha)
\ee
The upper limit in the above integral is $X$ for unbounded
distributions. But for bounded distributions, when correlations are
strong (large $B$), the limits in case (ii) described below
(\ref{Jsimple}) apply. On dividing the above equation by
$I_B(X;\alpha)$, it follows that the average walk length
(\ref{Jsimple}) can be written  as  
\be
{\bar J}_B (L|f_0)=B (\beta_\kappa \ln L_B + c_\kappa) -\frac{\beta_\kappa B}{\kappa}
\left( \frac{1}{\alpha-1}- \frac{\partial}{\partial \alpha} \frac{\ln
I_B(B f_0;\alpha)}{B} \right)\bigg|_{\alpha=1+\frac{1}{\kappa}} 
\label{Jfin}
\ee
Our task is now reduced to finding the sum distribution
  $I_B(X)$ for the 
  various EVT domains which we describe below.

\noindent{\it Weibull class:} According to the large deviation
principle, for large $B$, the distribution $I_B(X)$ is 
of the form \citep{Touchette:2009},  
\be
I_B(X \simeq B x) \sim e^{B r(x)} 
\label{ldt}
\ee
where the rate function $r(x)$ can be determined as described below. 
On using the integral representation of the Dirac delta function in
(\ref{sumI}), we get    
\bea
I_B(X) &=& \frac{1}{2 \pi}\int_{-\infty}^\infty dk ~e^{i k X} \left(
\int_{-\infty}^\infty dy~ e^{-i k y} g(y) \right)^B \\
&=& \frac{1}{2 \pi i}\int_{-i \infty}^{i \infty} d \omega ~e^{B
  (\omega x+ \ln {\tilde g}(\omega))}
  \label{IB}
\eea
where ${\tilde g}(\omega)=\int_0^\infty dt ~g(t)~ e^{-\omega t}$ is 
the Laplace transform of the distribution function $g(t)$. 
Evaluating the RHS of (\ref{IB}) using the saddle point method for large
$B$ \citep{Bender:1999}, we get 
\be
\frac{\ln I_B(X)}{B}=r(x)=\omega_* x+ \ln {\tilde g}(\omega_*)
\label{rate}
\ee
where the saddle point $\omega_*$ is real and given by 
\be
{\frac{d \ln  {\tilde g}}{d \omega}}\bigg|_{\omega=\omega_*}=-x
\label{saddle}
\ee

\begin{figure}
\centering
\includegraphics[width=0.8 \linewidth,angle=270]{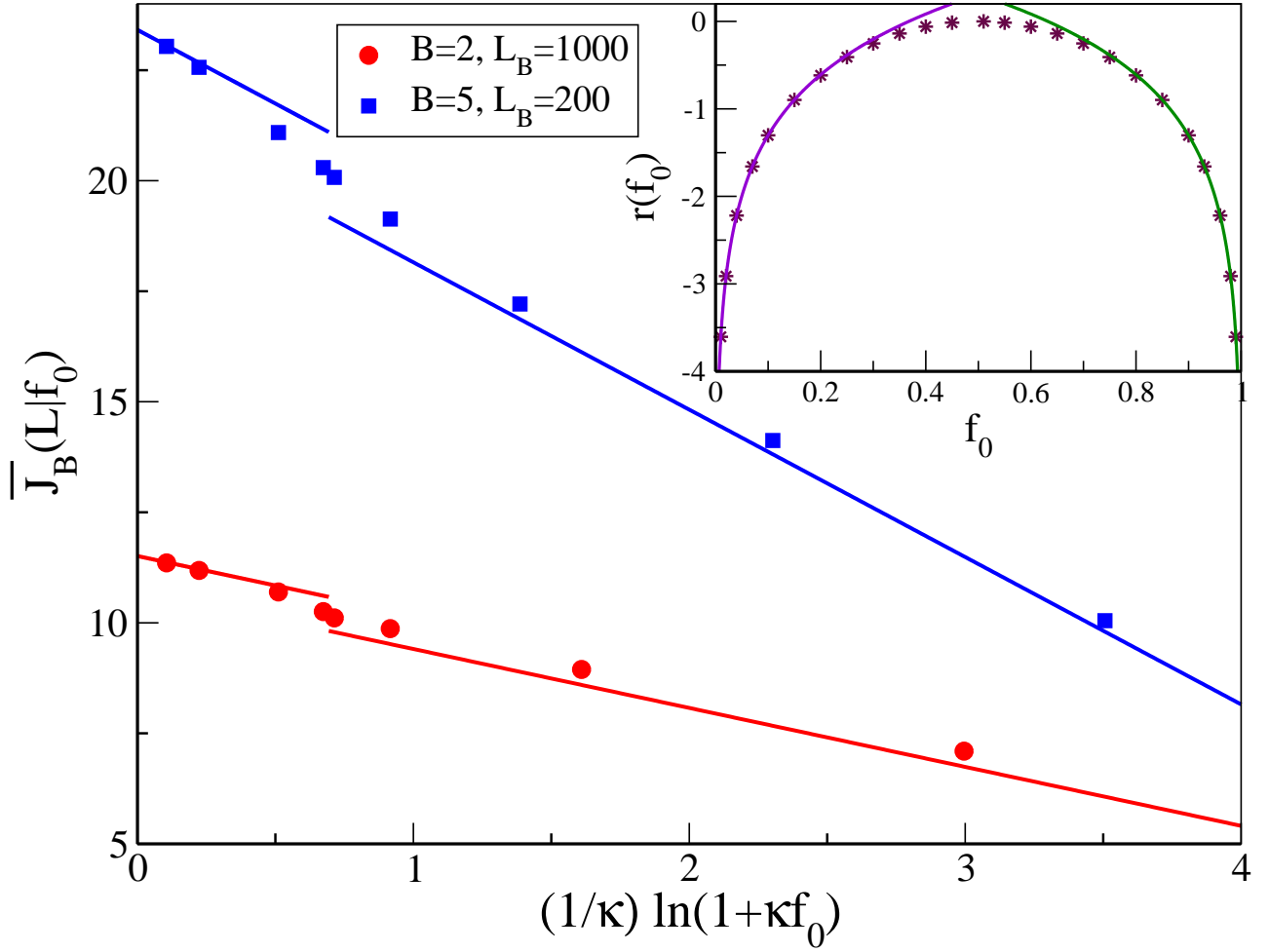}
\caption{(Color online) Main: Plot shows the variation of the average walk length with initial
  fitness for the linear model on correlated fitness landscapes for various $B$ when
  $\kappa=-1$. The
  theoretical predictions (\ref{uni_smallf02}) and (\ref{J_neg}) (lines)
  are compared against the  
  simulation data (points). Inset: Plot shows the rate function for
  $\kappa=-1$ 
  obtained using (\ref{rate}) and (\ref{star}) (points) and the analytical formulae 
(\ref{small}) and (\ref{large}).}
\label{uni_correl}
\end{figure}

The Laplace transform of the distribution $g(t)$  in (\ref{gt}) is given by
\be
{\tilde g}(\omega)=e^{\eta} \left[(\alpha-1)
E_{\alpha}(\eta) + \eta^{\alpha-1}
\Gamma(2-\alpha)  \right]
\ee
and the function $\omega_*(f_0)$ is a solution of the equation 
\be
{\cal T}(\omega_*)=\frac{(\alpha-1) \omega^2
  (E_{\alpha-1}(\eta)-E_\alpha(\eta))-\eta^\alpha \kappa^2 (\eta+
  \alpha-1) \Gamma(2-\alpha)}{\omega \kappa ((\alpha-1) \omega
  E_\alpha (\eta)+\eta^\alpha \kappa \Gamma(2-\alpha))}
\bigg|_{\omega=\omega_*}= f_0 
\label{star}
\ee
where $\eta=\omega/\kappa$, $E_\alpha(\eta)=\int_1^\infty dx~e^{-\eta
  x} x^{-\alpha}$ is the exponential integral and $\Gamma(n+1)=n!$ is the
gamma function. The function ${\cal T}(\omega_*)$ in the above equation decreases from its maximum value $-1/\kappa$ to zero as $\omega_*$ is increased from $-\infty$ to $\infty$. 
Using the asymptotic expansion of the exponential integral
\citep{Abramowitz:1964}, we find that 
\begin{subnumcases}
{{\cal T}(\omega_*)=}-\kappa^{-1}+ (1-\alpha) \omega_*^{-1} ~,~\omega_* \to -\infty
\label{neg} \\
\omega_*^{-1} ~,~ \omega_* \to \infty
\label{pos}
\end{subnumcases}
When the initial fitness is large (small), $f_0$ equals the left hand side (LHS) of
(\ref{star}) when $\omega_*$ is negative (positive). Then using 
(\ref{neg}) and (\ref{pos}) in (\ref{rate}), we find the rate function to be 
\begin{numcases}
{r(f_0) \approx} 1+\ln((\alpha-1) \kappa f_0)+ \ln (1-\alpha \kappa
f_0) &, $f_0 \ll {\cal T}(0)$ 
\label{small}\\
 1-\alpha- (1-\alpha) \ln \left( \frac{1-\alpha}{1+\kappa
  f_0}\right) +\ln (\Gamma(2-\alpha)) &, $f_0 \gg {\cal T}(0)$
  \label{large}
\end{numcases}
where ${\cal T}(0)=(1-\kappa)^{-1}$. 
The above expression for the rate function is
compared against the results from numerical simulations for uniformly
distributed fitnesses in the inset of Fig.~\ref{uni_correl}, and we
see a good agreement for $f_0 < 0.3$ and $> 0.7$. 
For small $f_0$, using (\ref{Jfin}), we obtain  
\bea
\frac{{\bar J}_B (L|f_0)}{B} &=& \beta_\kappa \ln L_B + c_\kappa- \frac{\beta_\kappa
  f_0}{1-(1+\kappa) f_0} \label{uni_smallf0} \\ 
&\approx & {\bar J} (L_B|f_0) \label{uni_smallf02}
\eea
while for large $f_0$,  we get 
\bea
\frac{{\bar J}_B (L|f_0)}{B} &=& \beta_\kappa \ln L_B + c_\kappa
-\frac{\beta_\kappa}{\kappa} (\kappa +\ln(-\kappa)+\ln(1+\kappa
f_0)+H_{-\frac{1}{\kappa}} -\gamma) \label{uni_largef0}\\
&=& {\bar J} (L_B|f_0) -\frac{\beta_\kappa}{\kappa} (\kappa +\ln(-\kappa)+H_{-\frac{1}{\kappa}} -\gamma)
\label{J_neg}
\eea
where the Euler-Mascheroni constant $\gamma \approx 0.577$. The walk
length expressions above can be succinctly written as 
\be
{\bar J}_B (L|f_0)={\bar J}_B (L|0) -\frac{B \beta_\kappa}{\kappa} \ln(1+\kappa
f_0)
\ee
and shows that the walk for nonzero fitness is shorter, as one
would intuitively expect. For $\kappa=-1$, the  
equations (\ref{uni_smallf0}) and (\ref{uni_largef0}) are compared
against the numerical results in Fig.~\ref{uni_correl}, and we see
that the theoretical prediction for the walk length matches the
simulation results quite 
well in the range of initial fitness values  where the rate function agrees.

\begin{figure}
\centering
\includegraphics[width=0.8 \linewidth,angle=270]{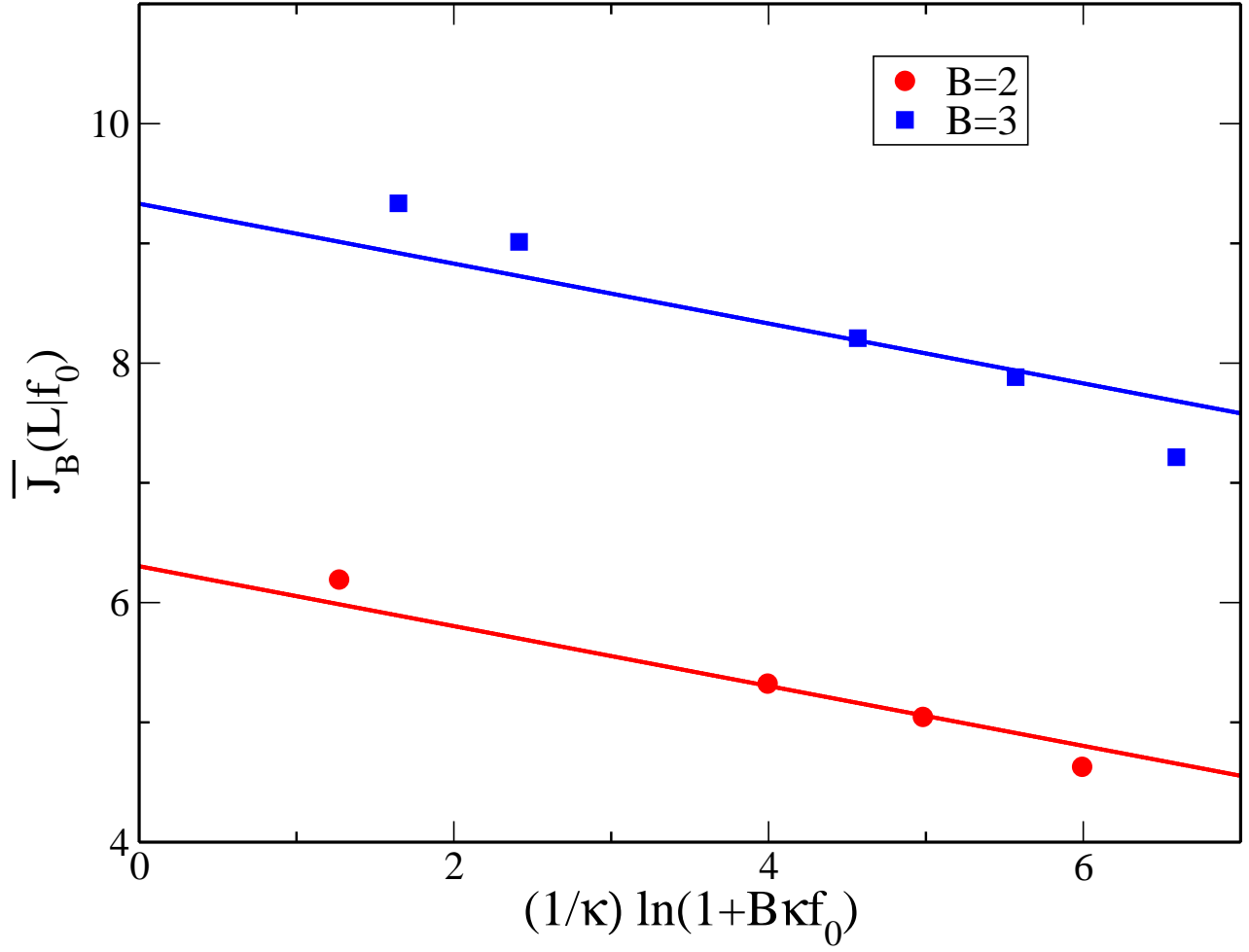}
%{L1000_uni_sim_B1_B2_B10_approx.ps} 
\caption{(Color online) Plot shows the variation of the average walk length with the initial
  fitness for the linear model on correlated fitness landscapes for various $B$ when
  $\kappa=2/3$ and $L_B=1000$. The  
  theoretical prediction (\ref{J_p}) (lines) is compared against the 
  simulation data (points).}
\label{k2by3_correl}
\end{figure}

\noindent{\it Fr{\'e}chet class:}  In this case, the sum
distribution (\ref{sumI}) for large $B f_0$ is given by \citep{Sornette:2000} 
\be
I_B(B f_0;\alpha) \sim B g(B f_0)
\label{kptail}
\ee
whose tail behavior is the same as that of the fitness distribution
$g(f)$.
Using this in (\ref{Jfin}), we
immediately find   
\bea
{\bar J}_B (L|f_0) &=& B (\beta_\kappa \ln L_B + c_\kappa)
-\frac{\beta_\kappa}{\kappa} (\ln (1+ B \kappa f_0)+ \kappa (B-1))  
\label{J_p}
\\
&\approx& B (\beta_\kappa \ln L_B + c_\kappa) -\frac{\beta_\kappa}{\kappa} (\ln (\kappa
f_0)+ \ln B + \kappa (B-1))  
\label{J_pos}
\eea
We note that the above answer matches with (\ref{B2Fre}) for the two
block model discussed in the last subsection. 
The above equation states that the average walk length decreases
logarithmically with initial fitness but, 
unlike in the Weibull and Gumbel domain, the coefficient of $\ln f_0$
does not scale with the number of blocks. Thus in this case
\be
{\bar J}_B (L|f_0)= {\bar J}_B (L|0) -\frac{\beta_\kappa}{\kappa}
\ln(1+B \kappa f_0)
\ee
In Fig.~\ref{k2by3_correl}, 
  the above expression is compared with the simulation
  data for $\kappa=2/3$, and we see a good quantitative agreement between the theory and the simulations.

%=============================================================================
%=============================================================================

\section{Walk length in the full model}
\label{full}

As mentioned in Sec.~\ref{aw_model}, the transition probability
(\ref{linT}) used to calculate the walk length is valid only when the
relative fitness difference is small. However, large fitness
differences during successive steps in the walk can occur when the
initial fitness is small or if the fitness distribution has a fat
tail \cite{Seetharaman:2014}. In such cases, the approximation
(\ref{linT}) breaks down, and we 
should consider the full transition probability 
(\ref{Tfull}). We have not been able to obtain analytical results for
this model, and present our simulation results below.  

As in the linear model, the walks are long for the full model when the
initial fitness is low or when the fitness are correlated 
\cite{Seetharaman:2014}. However qualitative difference between the 
linear and the full model is seen with regard to the walk length dependence on the
extreme value domain. As explained in Sec.~\ref{wl_uncorr}, the divergence of
the denominator on the RHS of (\ref{linT}) is responsible for the
independence of the walk length on the initial fitness when $\kappa > 1$
in the linear model. However the normalisation constant in (\ref{Tfull}) remains finite for all $\kappa$
and therefore the walk length always decreases with increasing $f_0$
here. The inset of Fig.~\ref{trans_fig} shows that the full model is
approximated very well by the linear model in the Weibull domain, and
is a reasonable approximation in the Gumbel domain. This agreement is
explained by the fact that the fitness difference between successive
steps are indeed small in these two domains as discussed in
\cite{Seetharaman:2014}. However in the Fr{\'e}chet domain, the
relative fitness differences between the successive steps in the
adaptive walk can be as large as hundred \cite{Seetharaman:2014} thus
rendering the linear model invalid. For a fixed initial fitness rank,
the inset of Fig.~\ref{trans_fig} shows that in the full model, the
walk length increases 
with increasing $\kappa$ in the Fr{\'e}chet domain. Thus the behaviour
of the walk length is nonmonotonic in $\kappa$ with the minimum  occurring
in the Gumbel domain.

\begin{figure}
\centering
\includegraphics[width=0.8
 \linewidth,angle=270]{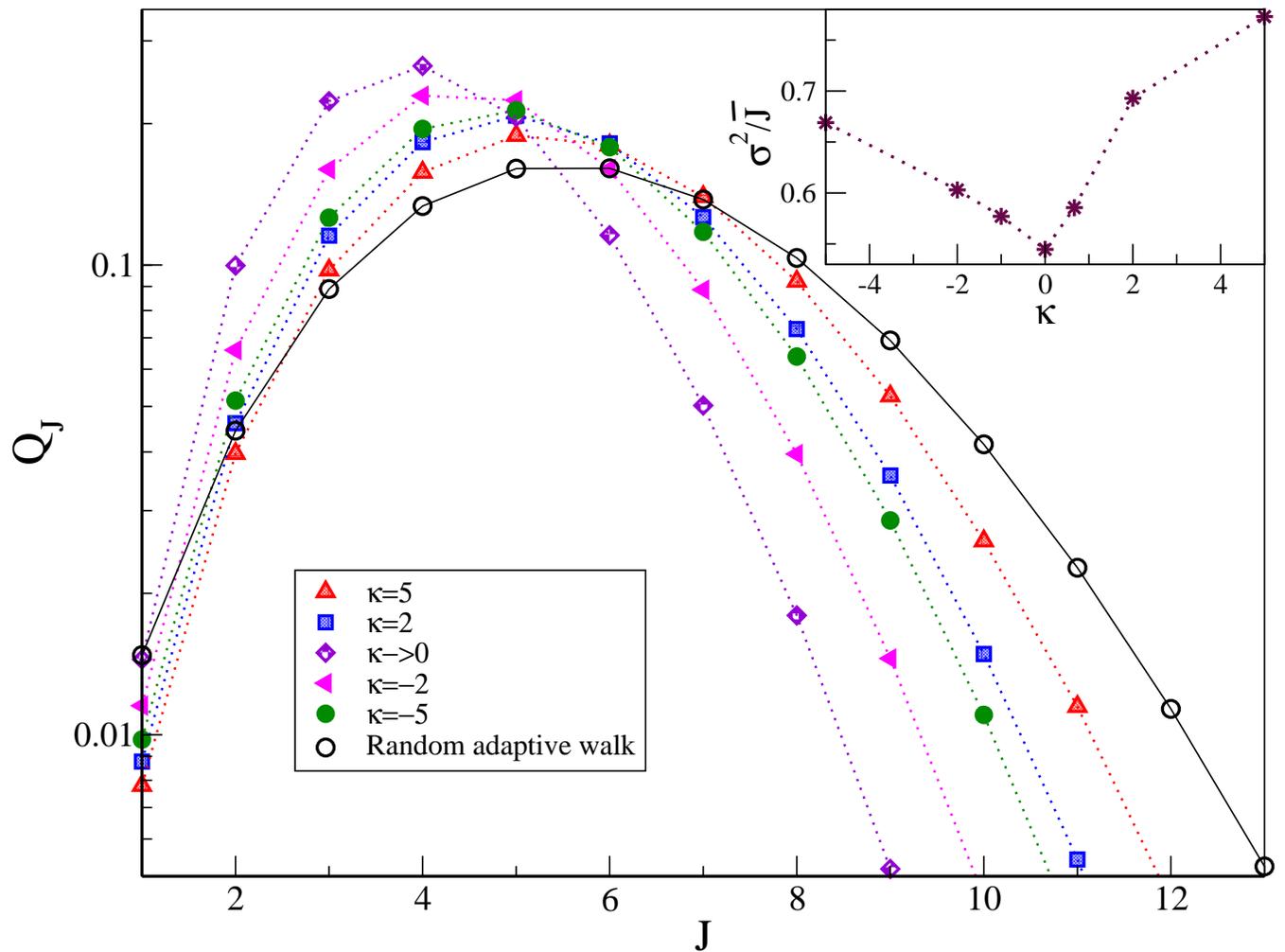} 
\caption{(Color online) Plot to show the simulation data for the walk length
  distribution (main) and the index of dispersion (inset) of
  the walk length for various $\kappa$ when $L=1000$ using the
  full model on uncorrelated fitness landscapes. In the inset,
  $({1}/{\kappa}) \ln(1+\kappa f_0)=2$.}       
\label{B1_full}
\end{figure}

Figure \ref{B1_full} shows the distribution of the walk length for
various $\kappa$ and uncorrelated fitnesses, and we
observe that as $|\kappa|$ increases, this distribution approaches the
corresponding result for the random adaptive walk where the walk
distribution is known to be a Poisson 
distribution with mean $\ln L$ \cite{Flyvbjerg:1992}. 
A related quantity is the index of dispersion
of the walk length which is the ratio of the 
variance to the mean which is shown in the inset of Fig. \ref{B1_full} and
displays a nonmonotonic behaviour with the minimum 
occurring at $\kappa=0$ and approaching unity for $\kappa \to \pm
\infty$. A similar nonmonotonic behavior is seen in the linear model
but in that case, the 
index of dispersion approaches unity when $\kappa \to -\infty$ and one
\cite{Neidhart:2011}.

%=============================================================================
%=============================================================================

\section{Conclusion}
\label{discussion}

In this article, we studied a model of adaptation in which beneficial
mutations sweep the population sequentially as it adapts by climbing
up a rugged fitness landscape. The broad question addressed here is
regarding the average number of adaptive mutations that occur until
the population reaches a local fitness peak. This quantity has been
measured in recent experiments on various systems like bacteriophage {\it $\phi X174$} \cite{Rokyta:2009},
fungus {\it A. nidulans} \cite{Gifford:2011} and bacteria {\it E. coli}
\cite{Sousa:2012}. Theoretically, the number of adaptive changes have 
been calculated on  
uncorrelated fitness landscapes for zero initial fitness
\cite{Jain:2011d,Jain:2011e} and high initial rank
\cite{Gillespie:1983,Orr:2002,Neidhart:2011}. Some studies
for correlated fitnesses have also been carried out
\cite{Orr:2006a,Jain:2011d,Filho:2012,Neidhart:2014}. Here we have extended the
previous works 
and studied how the length of the adaptive walk depends on the initial 
fitness, extreme value domains and fitness correlations. 

For the linear model that assumes small relative fitness differences in all
the extreme value domains, we find that the walk length
decreases with increasing initial fitness logarithmically provided the
mean of the fitness distribution is finite, otherwise it remains a
constant. The walks are found to be shorter for fitness distributions
that decay slower - in the limit $\kappa \to \infty$, the walk
length approaches the greedy walk limit (\ref{GAW}) while in the other
extreme of $\kappa \to -\infty$, it tends to the random adaptive  walk
(\ref{RAW}) \cite{Joyce:2008}. The logarithmic variation with the same
dependence on the fitness distribution as here has also been seen in other
systems \cite{Neidhart:2011}. 
On correlated fitness landscapes, the previous studies have been
largely numerical \cite{Orr:2006a,Filho:2012,Neidhart:2014}, while here we have presented 
analytical results. Interestingly, the large deviation theory finds 
an application in the calculation of the walk length for correlated
fitnesses. We find that, as on uncorrelated fitness landscapes, the
walk length decreases with increasing initial fitness and
GPD exponent $\kappa$. But increasing fitness correlations also
lengthen the adaptive walk since the population encounters lesser
number of local fitness peaks. Our detailed analysis shows that the 
walk length difference ${\bar J}_B(L|f_0)-{\bar J}_B(L|0)$ 
scales linearly with the number of blocks (that are a
measure of correlations) in the Weibull and Gumbel domains, and shows a weaker logarithmic dependence on the number of blocks in the Fr{\'e}chet
domain. For the sake of completeness, we also performed simulations for $\kappa > 1$ and found that the average walk length in this case shows a linear dependence on the block number (data not shown). These results for the linear model are summarised in Table~\ref{tab}.
%The summary of our results in this part would be that, on uncorrelated fitness landscapes the average walk length has a logarithmic depends on the initial fitness rank as long as $\kappa <1$ and beyond that, it becomes independent of initial fitness. However, on correlated fitness landscapes while the average walk length has a linear dependence on the number of blocks in the Weibull and Gumbel domains, the dependence is logarithmic in the Fr{\'e}chet domain between $0< \kappa < 1$ \tc{blue} {and becomes linear beyond that. to be checked}
%}

For the full model that is not restricted to small relative fitness
differences, we find that the walk length decreases with initial
fitness for all $\kappa$, unlike in the linear model. The walk length
is however seen to match quantitatively well in the Weibull domain
where small fitness differences arise \cite{Seetharaman:2014}. In
contrast, in the Fr{\'e}chet domain, even the qualitative trends in
the two models are opposite: while the walk length decreases with
increasing $\kappa (< 1)$ in the linear model, it increases in the
full model. Thus in the full model, the walk is shortest in the
Gumbel domain. An analytical understanding of these results is however
not available. 

Experiments show that a moderately sized population reaches a fitness
plateau in two to four substitutions
\cite{Rokyta:2009,Gifford:2011,Sousa:2012} (although one population
has been seen to gain nine beneficial mutations as well
\cite{Rokyta:2009}) thus indicating that the adaptive walks are
generally short.  
An inverse relationship between the initial fitness and the walk length
has been observed in some experiments \cite{Rokyta:2009,Sousa:2012} in 
agreement with the full model. However a constant walk length
independent of initial fitness has been seen in a recent experiment
\citep{Gifford:2011}. As described above, the full model predicts the walk length to be a nonmonotonic function of the parameter $\kappa$. The adaptive walk is expected to last longer in the experimental set ups in which Weibull \cite{Rokyta:2009, Bataillon:2011} or Fr{\'e}chet \cite{Schenk:2012} domain is observed than in the ones in which the distribution of beneficial mutations has an exponential tail \cite{Sanjuan:2004a, Rokyta:2006, Kassen:2006, Maclean:2010}.  However the walk length has not been measured in these experiments, while in the walk length experiments \cite{Rokyta:2009,Gifford:2011,Sousa:2012}, the extreme value domain of the beneficial mutation has not been studied and therefore presently the theoretical predictions regarding the connection between the extreme value theory and the length of the adaptive walk remains experimentally untested. 
Although some of the available experimental results
are in qualitative  
agreement with the theoretical predictions described above, a quantitative
comparison between the experiments and the theory 
seems difficult.  This is because in experiments measuring the walk
length, the walk is assumed to terminate if the fitness remains 
constant over some time period but that need not imply that the
adaptation is over \citep{Rokyta:2009}. Besides most experiments
\citep{Gifford:2011} cannot measure mutations whose fitness difference   
is below a threshold value and miss out on
mutations conferring slight benefit thus underestimating the walk
length. A better understanding of the theoretical results vis-{\`a}-vis the
experimental ones remains a goal for the future. 

%On the theoretical front, our previous work \cite{Seetharaman:2014} has noted that the fitness difference between successive mutations can indicate the EVT domain of the fitness of beneficial mutations since this quantity increases, is a constant and decreases in the Weibull, Gumbel and Fr{\'e}chet domains respectively.  As an extension of that work, here we study how 
%the
%walk length depends on these three extreme value domains. 
%
%\tc{red}{Unlike uncorrelated fitness landscapes where the average walk length has a logarithmic dependence on the rank of the initial fitness due to the population seeing only the higher fitnesses corresponding to lower ranks over the course of the adaptive walk,  on correlated fitness landscapes the logarithmic dependence on the initial rank may not exist in all cases due to the independent evolution of each block. }

%=============================================================================
%=============================================================================
\appendix

\section{Solution of the generating function equation (\ref{G})}
\label{Agenfn}

The probability distribution ${\cal P}_J(f|f_0)$ obeys the following
recursion equation \citep{Jain:2011d} 
\be
{\cal P}_{J+1}(f|f_0)= \int_{f_0}^{f} dh ~T(f \leftarrow h)~(1-q^L(h))~
{\cal P}_J(h|f_0)~,~J \geq 0
\label{main}
\ee
where $T(f \leftarrow h)$ is given by (\ref{linT1}). The above equation simply means that the population
moves from fitness $h$ to a higher fitness $f$ at the next step with
probability $T(f \leftarrow h)$ provided at 
least one fitter  mutant is available, the probability of whose is given by 
$1-q^L(h)$. For monomorphic initial condition with fixed fitness
$f_0$, we have the boundary conditions  
\bea
{\cal P}_J(f|f_0)&=&\delta(f-f_0) \delta_{J,0} \label{bc1} \\
{\cal P}'_J(f_0|f_0)&=& T'(f \leftarrow f_0)~ p(f_0)~ (1-q^L(f_0)) \delta_{J,1}\label{bc2}
\label{mono}
\eea
Equation (\ref{bc1}) is self explanatory and (\ref{bc2}) is obtained
by applying (\ref{bc1}) on the first derivative of (\ref{main})
w.r.t. $f$ \citep{Jain:2011d}. For the linear model with transition probability (\ref{linT1}), the integral equation (\ref{main}) can be recast as a
second order differential equation (\ref{final}).

For infinitely long sequences, the cumulative probability
distribution $q^L(h)\rightarrow 0$ and the differential equation
(\ref{G}) for the generating function $G(x,f)$ reduces to  
\be
G''(x,f)=\dfrac{x (1-\kappa)}{(1+\kappa f)^2}G(x,f)
\label{diff}
\ee
From (\ref{bc1}) and (\ref{bc2}), we have 
\bea
G(x,f_0)&=&0 
\label{bg1} \\
G'(x,f_0)&=&\frac{x}{\int_{f_0}^u dg~(g-f_0) ~p(g)}
  \label{bg2}
\eea
The solution of (\ref{diff}) subject to above initial conditions is
given by \citep{Bender:1999} 
\bea
 G(x,f)=\dfrac{x (1-\kappa) (1+\kappa
 f_0)^{1/\kappa}}{\sqrt{\kappa^2+4x(1-\kappa)}}\left[\left( \dfrac{1+\kappa
 f}{1+ \kappa f_0}\right)^{\alpha_+}-\left( \dfrac{1+\kappa f}{1+
 \kappa f_0}\right)^{\alpha_-}\right]
\label{genfn}
\eea
where 
\be 
\alpha_{\pm}=\dfrac{1}{2}\left(1\pm\sqrt{1+\frac{4 x(1-\kappa)}{\kappa^2}}\right)
\ee

The functions $a_1, a_2$ appearing in (\ref{Hcases1}) and
  (\ref{Hcases2}) can be calculated explicitly using the above result.  
In terms of $z$ defined in (\ref{zdef}), the solution (\ref{genfn}) for $\kappa \neq 0$ can be written as 
\begin{equation}
 G(x,z)=\dfrac{x(1-\kappa)(1+\kappa
 f_0)^{1/\kappa}}{\sqrt{\kappa^2+4x(1-\kappa)}}\left[\left(z \dfrac{1+\kappa\tilde{f}}{1+\kappa f_0}\right)^{\alpha_+}-\left(z\dfrac{1+\kappa\tilde{f}}{1+\kappa f_0}\right)^{\alpha_-}\right]
\end{equation}
Comparing the above equation with (\ref{Gless}), we get 
\begin{subequations}
\label{a12_kne0}
\begin{align}
 a_1 = \dfrac{x (1-\kappa)(1+\kappa
 f_0)^{1/\kappa}}{\sqrt{\kappa^2+4x(1-\kappa)}}\left( \dfrac{1+\kappa\tilde{f}}{1+\kappa f_0}\right)^{\alpha_+}\\
a_2 = -\dfrac{x (1-\kappa)(1+\kappa
 f_0)^{1/\kappa}}{\sqrt{\kappa^2+4x(1-\kappa)}}\left( \dfrac{1+\kappa\tilde{f}}{1+\kappa f_0}\right)^{\alpha_-}
\end{align}
\end{subequations}
For exponentially distributed fitnesses, taking the limit $\kappa\rightarrow0$ in (\ref{genfn}) and using (\ref{zdef}), we find that
\bea
G(x,z)=\dfrac{\sqrt{x}e^{f_0}}{2}\left( e^{z \sqrt{x}} e^{({\tilde f}-f_0)\sqrt{x}}-e^{-z \sqrt{x}} e^{-({\tilde f}-f_0)\sqrt{x}}\right)
\eea
from which we obtain
\begin{subequations}
\label{a12_ke0}
\begin{align}
a_1= \dfrac{\sqrt{x}e^{f_0}}{2}e^{({\tilde f}-f_0)\sqrt{x}}\\
a_2= -\dfrac{\sqrt{x}e^{f_0}}{2}e^{-({\tilde f}-f_0)\sqrt{x}}
\end{align}
\end{subequations}
%-----------------------------------------------------------------------

\section{Walk length using Gaussian approximation for
  exponentially distributed fitnesses}
\label{gauss}

By virtue of central limit theorem \citep{Sornette:2000}, the
distribution ${\cal N}_B(X)$ of the sum of $B$ i.i.d. random variables
is given by
\be
{\cal N}_B(X) = \frac{1}{\sqrt{2 \pi B \sigma^2}} ~\exp\left[-\frac{(X-B
    {\bar f})^2}{2 B \sigma^2} \right]
\label{clt1}
\ee
provided the mean ${\bar f}$ and the variance $\sigma^2$ of the parent distribution 
$p(f)$ exist. Since the Gaussian distribution is a good approximation
to the exact distribution of the sum when $X \sim B {\bar f}
\pm \sqrt{2 B \sigma^2}$, we expect that it will provide a good
estimate of the walk length when $f \sim B (f_0-{\bar f}) \pm \sqrt{2 B
  \sigma^2}$ in the integrand in (\ref{correl_exp2}). With increasing
$B$, as the core of the distribution ${\cal N}_B(X)$ moves rightwards
while the factor $f e^{-f}$ in the integrand peaks around one, the
overlap is significant when $f_0 \sim 1 \mp \sqrt{2 \sigma^2/B}$. 
Thus the Gaussian approximation for the sum distribution is likely to work
well in the neighborhood of initial fitness one. This can be seen more
explicitly as follows: using (\ref{clt1}) in the integral appearing in
(\ref{correl_exp2}), we get  
\bea
I_{clt} &=& \int_{0}^{B f_0} df ~f e^{-f} {\cal N}_{B-1}(B f_0- f) \\
&=& \frac{a e^{a^2} e^{-2 a b}}{\sqrt{\pi}}~ \left[
  e^{-(a-b)^2}-e^{-4 a^2}+\sqrt{\pi} (a-b) (\textrm{erf}(a-b)-\textrm{erf}(2 a)) \right]
\eea
where $a=\sqrt{(B-1)/2}$ and $b=(B f_0- B+1)/(2 a)$. For large $B$,
using the asymptotic expansion of error function, we get 
\be
I_{clt} \approx \frac{a e^{a^2} e^{-2 a b} e^{-(a-b)^2}}{2 \sqrt{\pi} (a-b)^2}
\ee
The expression (\ref{correl_exp2}) for the average walk length then 
gives 
\bea
{\bar J}_B (L|f_0) &=& B (\beta_0 \ln L_B + c_0) -\beta_0 B
\frac{e^{-f_0(f_0-1)}}{(2-f_0)^2}  \\
& \approx & B (\beta_0 \ln L_B + c_0) -B \beta_0 f_0  ~,~f_0
\to 1 
\eea
Thus it is only when the initial fitness is close to unity, the
Gaussian approximation captures the linear relationship between ${\bar J}$ and
$f_0$ correctly. 

\clearpage

\begin{table}[ht]
\begin{tabular}[c]{|c|c|c|}
\hline
 EVT domain & Dependence on initial rank & Dependence on number of blocks \\
 \hline
 \hline
 Weibull, $\kappa < 0$& Logarithmic  & Linear  \\
 \hline
 Gumbel $\kappa \to 0$& Logarithmic & Linear  \\
 \hline
 Fr{\'e}chet, $0 < \kappa<1$& Logarithmic  & Logarithmic  \\
 Fr{\'e}chet, $\kappa>1$& {Independent}& Linear\\
 \hline
\end{tabular}
\caption{Table summarising the dependence of the walk length on extreme value domains, initial fitness and fitness correlations in the linear model.}
\label{tab}
\end{table}

\clearpage
%=============================================================================
%=============================================================================

\clearpage
%=============================================================================
%FIGURES
%=============================================================================

%=============================================================================

\end{document}